\documentclass{article}

\usepackage{eusipco2011}  
\usepackage{amssymb,amsmath,epsfig}
\usepackage{array} 
\usepackage{colortbl} 
\usepackage[table]{xcolor} 
\usepackage{multirow} 

\title{Assessment of Audio Features for Automatic Cough Detection}

\name {Thomas Drugman, Jerome Urbain, Thierry Dutoit}

\address{TCTS Lab, Faculty of Engineering, University of Mons\\
31, Boulevard Dolez - 7000 Mons - Belgium\\
phone: + (32) 65 37 47 49, fax: + (32) 65 37 47 29, email: thomas.drugman@umons.ac.be \\
web: http://tcts.fpms.ac.be/$\sim$drugman/ }

\begin{document}

\maketitle

\begin{abstract}
This paper addresses the issue of cough detection using only audio recordings, with the ultimate goal of quantifying and qualifying the degree of pathology for patients suffering from respiratory diseases, notably mucoviscidosis. A large set of audio features describing various aspects of the audio signal is proposed. These features are assessed in two steps. First, their intrisic potential and redundancy are evaluated using mutual information-based measures. Secondly, their efficiency is confirmed relying on three classifiers: Artificial Neural Network, Gaussian Mixture Model and Support Vector Machine. The influence of both the feature dimension and the classifier complexity are also investigated.
\end{abstract}

\section{Introduction}\label{sec:Intro}

For children as well as for adults, cough is in pneumology the commonest syndrom. It is a daily and very frequent reason of seaking advices to the general practitioner (around 20\% of consultations for children below 4 years old), the paediatrician and the pneumologist (for whom chronic cough represents one third of consultations). The impact of cough, notably chronic coughing, on life quality can be important.

The severity of cough can be evaluated by asking patients to fill in forms about their perception of the syndrom. However such a subjective assessment of cough has been shown \cite{CorrelQL} to be only slightly correlated to its objective characterization (using audio or video recordings for example). Medical literature on this topic therefore underlines the lack of a tool allowing the automatic, objective and reliable quantification of this symptom. This latter step is notably anterior to any correct evaluation of possible treatments.

Some approaches have been proposed to address the automatic detection of cough \cite{Smith}. These systems generally couple various sensors to the audio signal (see \cite{Smith} and references in it): air coupled microphones, accelerometer, lapel microphone, free field microphone, throat microphone or contact sensor. Although reported results are encouraging \cite{Smith}, there is currently no standardization and very few of these approaches led to a commercialization. In addition, following the patient in ambulatory and 24h-long conditions (while preserving his daily habits) remains an open problem.

As a result, cough quantification in the majority of hospitals is still nowadays performed by a tedious task of manual counting from audio recordings, or for validation by comparison using simultaneous video recordings.

This paper focuses on the automatic detection of cough using only the audio signal, as a preliminary and necessary study for its further integration within a multimodal system. On an acoustic point of view, cough is described as a forced expulsive manoeuvre against a closed glottis that is associated with a characteristic sound \cite{ERS}. The main difficulty in detecting cough from audio recordings lies in its efficient discrimination with other audio non-cough events such as speech, laugh, or ambient noise.

The goal of this paper is to study which audio features and classifier are the most suited for automatic cough detection. For this, it is structured as follows. Section \ref{sec:Feat} proposes a large set of possible audio features for this purpose. The experimental protocol used for assessing these features is described in Section \ref{sec:Protocol}. Section \ref{sec:MI} first evaluates their significance using mutual information-based measures. These features are then integrated within three classifiers in Section \ref{sec:Classifier}: Artificial Neural Network (ANN), Gaussian Mixture Model (GMM) and Support Vector Machine (SVM). Finally Section \ref{sec:Conclu} concludes the paper.

\section{Audio Features for Cough Detection}\label{sec:Feat}

The various audio features that are used throughout this study are briefly presented in the following. These features can be divided into three categories: features describing the spectral contents, measures of noise, and prosody-related features. In our experiments, we also added the first and second derivatives for each of these features in order to integrate the sound dynamics. This leads to a total set of 105 descriptors whose relevance will be assessed in Sections \ref{sec:MI} and \ref{sec:Classifier}.  

\subsection{Features Describing the Spectral Contents}\label{ssec:Spectral}
Several features characterizing the spectral shape have been proposed in \cite{Peeters}. For a comprehensive description of the magnitude spectrum, the well-known \emph{Mel Frequency Cepstral Coefficients} (MFCCs, \cite{MFCC}) are extracted. 13 MFCCs (including the $0^{th}$ coefficient) are used to represent the spectral distribution within 13 perceptual sub-bands. Besides, several parameters describing the spectral shape are also employed. The \emph{Spectral Centroid} is defined as the barycenter of the amplitude spectrum. Similarly, the \emph{Spectral Spread} is the dispersion of the spectrum around its mean value. The \emph{Spectral Decrease} is a perceptual measure quantifying the amount of decreasing of the spectral amplitude \cite{Peeters}. Finally, the \emph{Spectral Variation} and \emph{Spectral Flux} characterize the amount of variations of spectrum along time and are based on the normalized cross-correlation between two successive amplitude spectra \cite{Peeters}. 

\subsection{Measures of Noise}\label{ssec:Noise}
Quantifying the level of noise in the audio signal is of interest for describing the cough sound. For this purpose, several measures are here suggested. First, the \emph{Harmonic to Noise Ratio} (HNR) is calculated for the frequency ranges [0-0.5kHz], [0-1.5kHz], [0-2.5kHz] and [0-3.5kHz] using the Voice Sauce toolkit freely available at: \emph{http://www.ee.ucla.edu/$\sim$spapl/voicesauce/index.html}. These latter parameters are respectively denoted \emph{HNR05}, \emph{HNR15}, \emph{HNR25} and \emph{HNR35} in the remainder of this paper. The \emph{Cepstral Peak Prominence} (CPP) is used as it has been shown to be correlated with the degree of breathiness in voice \cite{Shue}. The \emph{Spectral Flatness} measures the noisiness/sinusoidality of a spectrum (or a part of it). As suggested in \cite{Peeters}, we here calculate the spectral flatness in the four following frequency bands: [0.25-0.5kHz], [0.5-1kHz], [1-2kHz] and [2-4kHz]. The \emph{Zero-Crossing Rate} quantifies the number of times the signal crosses the zero axis. It is expected that the greater the amount of noise, the higher the amount of zero-crossing. As a last parameter quantifying the amount of noise in the audio signal, the \emph{Chirp Group Delay} (chirp GD) is a phase-based measure proposed in \cite{DrugmanPhase} for highlighting turbulences during glottal production.

\subsection{Prosody-related Features}\label{ssec:Proso}
In speech processing, prosody refers to the rhythm, stress and intonation of speech. It is generally reflected by clues such as volume, pitch and duration. We therefore use measures of energy and loudness which basically are informative mainly about the presence of audio activity. As it is known \cite{ERS} that for a three-phase cough sound, the last phase presents voicing, the fundamental frequency is estimated using the STRAIGHT technique \cite{STRAIGHT_F0}.

\section{Experimental Protocol}\label{sec:Protocol}

The database consists of audio signals captured by a cheap standard MP3 recorder in an hospital context. They were kindly provided by the belgian mucoviscidosis center at the Cliniques Universitaires Saint-Luc. Subjects are patients suffering from mucoviscidosis who had to spend a night at the hospital. The recorder was placed on their bedside table during the evening. Recordings then contain parasitical signals such as talking, laughing and TV, music or other types of noise, which can be confusing for detecting cough. The database is made of 5 minute-long recordings from 9 different patients, manually labeled in cough and non-cough segments.

Audio signals were downsampled from 44.1 kHz to 16 kHz. Features introduced in Section \ref{sec:Feat} were extracted every 10 ms on Hanning windows whose length is 25 ms. The relevance of these features is assessed in Sections \ref{sec:MI} and \ref{sec:Classifier}. First, an evaluation based on the Mutual Information (MI) is led in Section \ref{sec:MI}. This approach is advantageous as it is independent of any classifier. A method of feature selection based on MI is employed to reduce dimensionality. In a second step, these features are assessed in Section \ref{sec:Classifier} by being integrated within three classifiers: ANN, GMM and SVM.

\section{Mutual Information-based Assessment and Feature Selection}\label{sec:MI}

\subsection{Background on Mutual Information}\label{ssec:BackgroundMI}

The problem of automatic classification consists in finding a set of features $X_i$ such that the uncertainty on the determination of classes $C$ is reduced as much as possible \cite{FSBook}. For this, Information Theory \cite{Cover} allows to assess the relevance of features for a given classification problem, by making use of the following measures (where $p(.)$ denotes a probability density function):

\begin{itemize}
\item The entropy of classes $C$ is expressed as:
\begin{equation}
H(C)=-\sum_c{p(c)\log_2p(c)}
\label{eq:entropy}
\end{equation}
and can be interpreted as the amount of uncertainty on their determination.

\item The mutual information between one feature $X_i$ and classes $C$:
\begin{equation}
I(X_i;C)=\sum_{x_i}{\sum_c{p(x_i,c)\log_2\frac{p(x_i,c)}{p(x_i)p(c)}}}
\label{eq:MI}
\end{equation}
can be viewed as the information the feature $X_i$ conveys about the considered classification problem, i.e. the discrimination power of one individual feature.

\item The joint mutual information between two features $X_i$, $X_j$, and classes $C$ can be expressed as:
\begin{equation}
I(X_i,X_j;C)=I(X_i;C)+I(X_j;C)-I(X_i;X_j;C)
\label{eq:jointMI}
\end{equation}
and corresponds to the information that features $X_i$ and $X_j$, when \emph{used together}, bring to the classification problem. The last term can be written as:
\begin{equation}
\begin{split} I(X_i;&X_j;C)= \\
\sum_{x_i}\sum_{x_j}\sum_cp(x_i,x_j,c)\cdot&\log_2\frac{p(x_i,x_j)p(x_i,c)p(x_j,c)}{p(x_i,x_j,c)p(x_i)p(x_j)p(c)}
\end{split}
\label{eq:redundancy}
\end{equation}

An important remark has to be underlined about the sign of this term. It can be noticed from Equation \ref{eq:jointMI} that a positive value of $I(X_i;X_j;C)$ implies some \textbf{redundancy} between the features, while a negative value means that features present some \textbf{synergy} (depending on whether their association brings respectively less or more than the addition of their own individual information).

\end{itemize}

\subsection{Mutual Information-based Assessment}\label{ssec:AssessMI}

To evaluate the significance of the audio features proposed in Section \ref{sec:Feat}, the following measures are computed:

\begin{itemize}
\item the \textit{relative intrinsic information} of one individual feature $\frac{I(X_i;C)}{H(C)}$, i.e. the percentage of relevant information conveyed by the feature $X_i$,
\item the \textit{relative redundancy} between two features $\frac{I(X_i;X_j;C)}{H(C)}$, i.e. the percentage of their common relevant information,
\item the \textbf{relative joint information} of two features $\frac{I(X_i,X_j;C)}{H(C)}$, i.e. the percentage of relevant information they convey together.
\end{itemize}

For this, Equations \ref{eq:entropy} to \ref{eq:redundancy} are calculated. Probability density functions are estimated by a histogram approach. The number of bins is set to 50 for each feature dimension, which results in a trade-off between an adequately high number for an accurate estimation, while keeping sufficient samples per bin. Class labels correspond to the presence or not of a cough event.

\begin{table*}[htbp]
    \centering    
{\begin{tabular}{c|c|c|c|c|c|c|c|c|c|c|c|c|c|c|}
 & \rotatebox{90}{MFCC 0} & \rotatebox{90}{Zero-Crossing} & \rotatebox{90}{HNR05} & \rotatebox{90}{MFCC 1} & \rotatebox{90}{Chirp GD} & \rotatebox{90}{MFCC 4} & \rotatebox{90}{MFCC 3} & \rotatebox{90}{MFCC 8} & \rotatebox{90}{F0} & \rotatebox{90}{HNR15} & \rotatebox{90}{MFCC 5} & \rotatebox{90}{MFCC 2} & \rotatebox{90}{MFCC 6} & \rotatebox{90}{Flatness 2-4}\\
\hline
 MFCC 0 & \cellcolor{yellow}\textbf{47.2} & \cellcolor{green}59.0 & \cellcolor{green}56.3 & \cellcolor{green}57.5 & \cellcolor{green}55.4 & \cellcolor{green}54.7 & \cellcolor{green}54.2 & \cellcolor{green}53.6 & \cellcolor{green}54.2 & \cellcolor{green}54.0 & \cellcolor{green}53.0 & \cellcolor{green}53.0 & \cellcolor{green}52.9 & \cellcolor{green}52.9\\
\cline{2-15}
\hline
 Zero-Crossing & \cellcolor{orange}15.1 & \cellcolor{yellow}\textbf{26.8} & \cellcolor{green}41.8 & \cellcolor{green}35.0 & \cellcolor{green}38.0 & \cellcolor{green}41.8 & \cellcolor{green}40.4 & \cellcolor{green}38.9 & \cellcolor{green}36.1& \cellcolor{green}50.0& \cellcolor{green}40.2& \cellcolor{green}37.0& \cellcolor{green}39.1& \cellcolor{green}38.0\\
\cline{2-15}
\hline
 HNR05 & \cellcolor{orange}7.6 & \cellcolor{orange}1.8 & \cellcolor{yellow}\textbf{15.9} & \cellcolor{green}45.0 & \cellcolor{green}37.6 & \cellcolor{green}35.9 & \cellcolor{green}28.1& \cellcolor{green}29.8 & \cellcolor{green}32.7& \cellcolor{green}37.9& \cellcolor{green}32.1& \cellcolor{green}30.4& \cellcolor{green}30.2 & \cellcolor{green}28.0\\
\cline{2-15}
\hline
 MFCC 1 & \cellcolor{orange}20.5 & \cellcolor{orange}22.6 & \cellcolor{orange}2.6 & \cellcolor{yellow}\textbf{30.8} & \cellcolor{green}39.2 & \cellcolor{green}43.4 & \cellcolor{green}43.3 & \cellcolor{green}39.9 & \cellcolor{green}37.4& \cellcolor{green}51.0 & \cellcolor{green}41.0& \cellcolor{green}36.5& \cellcolor{green}40.4& \cellcolor{green}39.0\\
\cline{2-15}
\hline
 Chirp GD & \cellcolor{orange}18.0 & \cellcolor{orange}15.0 & \cellcolor{orange}5.4 & \cellcolor{orange}17.8 & \cellcolor{yellow}\textbf{26.2} & \cellcolor{green}43.7 & \cellcolor{green}42.9 & \cellcolor{green}35.2& \cellcolor{green}33.0& \cellcolor{green}46.1& \cellcolor{green}38.7& \cellcolor{green}32.2& \cellcolor{green}37.1& \cellcolor{green}31.8\\
\cline{2-15}
\hline
 MFCC 4 & \cellcolor{orange}16.8 & \cellcolor{orange}9.2 & \cellcolor{orange}5.2 & \cellcolor{orange}11.7 & \cellcolor{orange}6.7 & \cellcolor{yellow}\textbf{24.2} & \cellcolor{green}33.2& \cellcolor{green}38.5& \cellcolor{green}32.6& \cellcolor{green}42.8& \cellcolor{green}35.7& \cellcolor{green}35.0& \cellcolor{green}38.7& \cellcolor{green}37.3\\
\cline{2-15}
\hline
 MFCC 3 & \cellcolor{orange}4.1 & \cellcolor{orange}-2.5 & \cellcolor{orange}0.1 & \cellcolor{orange}-1.4 & \cellcolor{orange}-5.6 & \cellcolor{orange}2.1 & \cellcolor{yellow}\textbf{11.1}& \cellcolor{green}26.5& \cellcolor{green}20.5& \cellcolor{green}37.1& \cellcolor{green}33.4& \cellcolor{green}24.2& \cellcolor{green}27.2& \cellcolor{green}26.7\\
\hline
 MFCC 8 & \cellcolor{orange}7.4 & \cellcolor{orange}1.7  & \cellcolor{orange} 0.9 & \cellcolor{orange}4.7& \cellcolor{orange}4.8 & \cellcolor{orange}-0.5 & \cellcolor{orange}-1.6& \cellcolor{yellow}\textbf{13.8}& \cellcolor{green}22.9& \cellcolor{green}38.9& \cellcolor{green}34.9& \cellcolor{green}26.3& \cellcolor{green}29.3& \cellcolor{green}25.2\\
\cline{2-15}
\hline
 F0 & \cellcolor{orange}-1.6 & \cellcolor{orange}-3.9 & \cellcolor{orange}-11.1 & \cellcolor{orange}-1.3 & \cellcolor{orange}-1.4 & \cellcolor{orange}-3.0 & \cellcolor{orange}-3.9& \cellcolor{orange}-3.6 & \cellcolor{yellow}\textbf{5.0}& \cellcolor{green}45.4& \cellcolor{green}27.6& \cellcolor{green}21.8& \cellcolor{green}24.0& \cellcolor{green}21.1\\
\cline{2-15}
\hline
 HNR15 & \cellcolor{orange}22.9 & \cellcolor{orange}6.7 & \cellcolor{orange}6.9& \cellcolor{orange}9.7& \cellcolor{orange}10.0& \cellcolor{orange}11.4& \cellcolor{orange}4.2& \cellcolor{orange}4.9& \cellcolor{orange}-10.7& \cellcolor{yellow}\textbf{29.0}& \cellcolor{green}41.5& \cellcolor{green}37.7& \cellcolor{green}40.3& \cellcolor{green}38.1\\
\cline{2-15}
\hline
 MFCC 5 & \cellcolor{orange}15.3 &\cellcolor{orange}7.7 & \cellcolor{orange}5.9& \cellcolor{orange}10.9& \cellcolor{orange}8.6& \cellcolor{orange}9.9& \cellcolor{orange}-1.3 & \cellcolor{orange}0.0& \cellcolor{orange}-1.1& \cellcolor{orange}9.7& \cellcolor{yellow}\textbf{21.1}& \cellcolor{green}30.1& \cellcolor{green}29.6& \cellcolor{green}30.4\\
\cline{2-15}
\hline
 MFCC 2 & \cellcolor{orange}7.7&\cellcolor{orange}3.3 & \cellcolor{orange}0.1 & \cellcolor{orange}7.8 & \cellcolor{orange}7.5 & \cellcolor{orange}2.7 & \cellcolor{orange}0.4& \cellcolor{orange}0.9 & \cellcolor{orange}-2.9 &\cellcolor{orange}5.9 &\cellcolor{orange}4.4 &\cellcolor{yellow}\textbf{13.5} & \cellcolor{green}28.2& \cellcolor{green}24.9\\
\cline{2-15}
\hline
 MFCC 6 & \cellcolor{orange}10.6& \cellcolor{orange}3.9 & \cellcolor{orange}3.1& \cellcolor{orange}6.6& \cellcolor{orange}5.3& \cellcolor{orange}1.7& \cellcolor{orange}0.1& \cellcolor{orange}0.7& \cellcolor{orange}-2.3& \cellcolor{orange}6.1& \cellcolor{orange}7.8& \cellcolor{orange}1.5& \cellcolor{yellow}\textbf{16.2}& \cellcolor{green}27.1\\
\cline{2-15}
\hline
 Flatness 2-4 & \cellcolor{orange}7.3& \cellcolor{orange}1.8 & \cellcolor{orange}2.0& \cellcolor{orange}4.8& \cellcolor{orange}7.4& \cellcolor{orange}-0.1& \cellcolor{orange}-2.6& \cellcolor{orange}1.6& \cellcolor{orange}-2.7& \cellcolor{orange}4.9& \cellcolor{orange}3.6& \cellcolor{orange}1.5& \cellcolor{orange}2.1& \cellcolor{yellow}\textbf{13.0}\\
\cline{2-15}
\hline 
\end{tabular}}
\caption{Mutual information-based measures for the 14 first selected features (respecting the ranking). \emph{On the diagonal}: the relative intrinsic information. \emph{In the bottom-left part}: the relative redundancy between the two considered features. \emph{In the top-right part}: the relative joint information of the two considered features.}
\label{tab:MITrainingSet}
\end{table*}

Since 105 audio features were extracted, an exhaustive presentation of results cannot be detailed here. For the sake of clarity, Table \ref{tab:MITrainingSet} displays the MI-based values for the 14 first features (respecting the ranking) selected by the algorithm that will be described in Section \ref{ssec:FeatSelect}. The diagonal indicates the percentage of relevant information conveyed by each feature. It is worth noting that the selection technique accounts for the redundancy and synergy between features. Selected features are therefore not necessarily the ones presenting the highest individual discrimination power. In our results, we observed that features conveying the greatest relative intrinsic information are: MFCC 0 (47.25\%), the loudness (46.56\%), a measure of energy (39.88\%), HNR35 (38.74\%) and HNR25 (35.41\%). The three first features are related to the signal energy and are particularly informative about the presence of an audio event. Although individually interesting, these features are strongly redundant, with e.g a value of 41.63\% of relative redundancy between MFCC 0 and the loudness. A strong redundancy (30.84\%) is also observed between HNR35 and HNR25. The algorithm of feature selection presented in Section \ref{ssec:FeatSelect} therefore tends to give priority to slightly redundant (or even synergic) features.

The top-right part of Table \ref{tab:MITrainingSet} contains the values of relative joint information of two features, while the bottom-left part shows the relative redundancy between two features. The best combination of two features is MFCC 0 with the zero-crossing rate, bringing together 59\% of relative joint information. Inspecting the values of redundancy, it is worth observing that F0 extracted with STRAIGHT is synergic with all 13 other features. The set of 14 features is relatively weakly redundant, with a maximum relative redundancy of 22.9\% between MFCC 0 and HNR15, and a maximum synergy value of -11.1\% between F0 and HNR05. Note the absence of first or second derivative features in the selected subset.

\subsection{Mutual Information-based Feature Selection}\label{ssec:FeatSelect}
Several techniques of feature selection have been proposed in the literature \cite{FSBook}. An important category of such methods is the approach relying on mutual information \cite{Drugman-MI}. Computing MI from data requires the estimation of probability densities, which cannot be accurately done in high dimensions. This is why a majority of feature selection algorithms use measures based on up to three variables (two features plus the class label). Therefore, various MI-based strategies for feature selection have been proposed, all trying to deal with the issue of redundancy management. In this paper, we use the following algorithm which is known \cite{Drugman-MI} to provide among the best results. Let us denote $F$=\{$X_1$,$X_2$,...,$X_N$\} the initial set of $N$ features, and $S_k$ the selected subset (with $S_k \subseteq F$) of $k$ features at step $k$. The method is a greedy algorihm which starts from an empty set and which selects at each step $k$ the feature $Y_k$ maximizing:

\begin{equation}
Y_k = \arg\max_{X_i\in F\setminus S_{k-1}} [I(X_i;C)- \max_{Y_j\in S_{k-1}} I(X_i;Y_j;C)]
\label{eq:FS}
\end{equation}

considering that the redundancy between $X_i$ and the selected subset $S_{k-1}$ is dominated by the most redundant feature in it.

It is confirmed in Table \ref{tab:MITrainingSet} that selected features exhibit weak redundancy values, as it is penalized via the term in $I(X_i;Y_j;C)$ in Equation \ref{eq:FS}. It is also interesting to note that selected features arise from the three categories: prosody-related characteristics (MFCC 0 and F0), noise measures (zero-crossing rate, HNR05, chirp GD, HNR15 and flatness 2-4), as well as spectral-based parameters (MFCC 1 to 8). Since these features arise from complementary sources of information, it can be expected that redundacy has been appropriately taken into account.

\section{Classifier-based Assessment}\label{sec:Classifier}
The use of three types of classifiers is here investigated: ANN, GMM and SVM. We rely on Matlab implementations for ANN and GMM, and on the Torch toolbox \cite{Torch} for SVM. Evaluation is achieved using a 10-fold cross validation framework. This means that training is led on $90\%$ of the database (randomly chosen), and the $10\%$ remaining are used for the test. This operation is repeated 10 times (with exclusive subsets for testing), so as to cover the whole database for the evaluation. The system is then generally assessed through its averaged error rate. However, given that the database is strongly unbalanced, i.e the proportion of cough events (compared to non-cough) is highly under-represented, we preferred to rely on Receiver Operating Characteristic (ROC) curves. A ROC curve shows the True Positive Rate (TPR, or sensitivity) as a function of the False Positive Rate (FPR, or 1-specificity) as a discrimination threshold $\theta$ is varied. As a single measure of performance of the ROC curve, we defined the Revised Error Rate (RER) as:

\begin{equation}\label{eq:RER}
RER= \min_{\theta} \sqrt{(1-TPR(\theta))^2+FPR(\theta)^2}
\end{equation}

Indeed, an ideal classifier being characterized by a $TPR=100\%$ and a $FPR=0\%$, a single measure of performance is the Euclidian distance from the top-left corner to the ROC curve. As a consequence, the lower RER, the better the system. This criterion implies that an equal importance is given to both TPR and FPR. Based on a medical advice, TPR or FPR could be emphasized by weighting its importance in Equation \ref{eq:RER}.

\subsection{ANN-based Classification}\label{ssec:ANN}
An Artificial Neural Network (ANN) is a method of classification using an interconnected group of artificial neurons, and which allows a non-linear statistical modeling of the class posterior. It is here used for its ability to model complex relationships between inputs (audio features) and outputs (posterior probability of belonging to a given class).

\begin{figure}[!ht]
  \centering
  \includegraphics[width=0.45\textwidth]{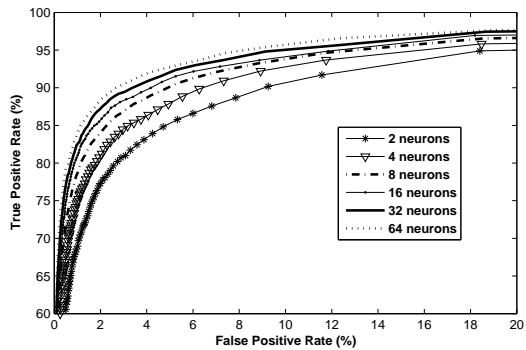}
  \caption{ROC curves obtained with the ANN classifier using 20 features and various numbers of neurons in the hidden layer.}
  \label{fig:ROC_ANN_Neur}    
\end{figure}

Our ANN implementation relies on the Matlab Neural Network toolbox. The ANN is made of a single hidden layer with a variable number of neurons whose activation function is an hyperbolic tangent sigmoid transfer function. Figure \ref{fig:ROC_ANN_Neur} displays the evolution of the ROC curves as a function of the number of neurons using the 20 first selected features. It is observed that the performance increases with the number of neurons. For information, a RER of 13.5\% is achieved with 2 neurons, 9.26\% with 32 and 8.78\% with 64 neurons.

\begin{figure}[!ht]
  \centering
  \includegraphics[width=0.45\textwidth]{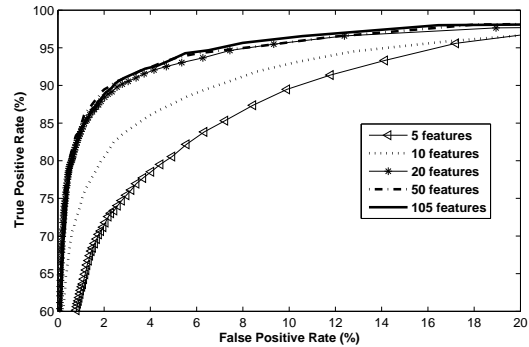}
  \caption{ROC curves obtained with the ANN classifier using 64 neurons in the hidden layer and various numbers of features.}
  \label{fig:ROC_ANN_Feat}    
\end{figure}

The impact of the number of features on the classifier performance is illustrated in Figure \ref{fig:ROC_ANN_Feat}, using 64 neurons in the hidden layer. Performance with 5 or 10 features is largely under what is obtained with more than 20 features. However, ROC curves carried out with 20, 50 and 105 features are very close, with respective RERs of 8.78\%, 8.13\% and 7.94\%. In other words, thanks to the efficient feature selection algorithm described in Section \ref{ssec:FeatSelect}, using only 20 features gives similar results to what is reached with 105 features, allowing an important dimensionality reduction.

For the best ANN configuration (64 neurons with 105 features), the following performance measures are obtained: \textbf{TPR=94.27\%, FPR=5.50\% and RER=7.94\%}.

\subsection{GMM-based Classification}\label{ssec:GMM}
\begin{figure}[!ht]
  \centering
  \includegraphics[width=0.45\textwidth]{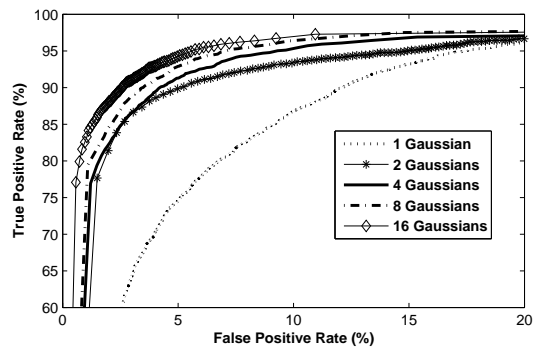}
  \caption{ROC curves obtained with the GMM classifier using 20 features and various numbers of Gaussians in the mixture.}
  \label{fig:ROC_GMM_Gauss}    
\end{figure}

A Gaussian Mixture Model (GMM) is a technique of classification in which the conditional probability for each class is approximated by a mixture of Gaussian distributions. In our Matlab implementation, GMMs are first initialized by a K-Means clustering step. The same number of Gaussians is used to model each class. Figure \ref{fig:ROC_GMM_Gauss} plots the ROC curves using 20 features and various numbers of Gaussians in the mixture. It is observed that cough detection gets better with an increasing number of Gaussians. 

\begin{figure}[!ht]
  \centering
  \includegraphics[width=0.45\textwidth]{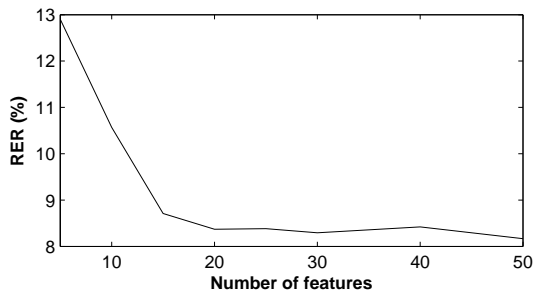}
  \caption{Evolution of RER as the number of features increases for a GMM classifier with 8 Gaussians.}
  \label{fig:RER_GMM}    
\end{figure}

In order to illustrate the influence of the feature dimension on the system, Figure \ref{fig:RER_GMM} displays the evolution of RER as a function of the number of features using 8 Gaussians. As it was the case for the ANN classifier, it turns out that using 20 features gives among the best results, and that the contribution when considering more features is minor.

For the best GMM configuration (16 Gaussians with 20 features), the following performance measures are obtained: \textbf{TPR=95.20\%, FPR=5.73\% and RER=7.48\%}. First, it is interesting to note that for audio-based cough detection GMM outperforms ANN, with a reduction of 0.46\% of RER. Secondly, it is worth emphasizing that only 20 features were used to reach that performance. For the same feature dimension, ANN obtains a RER of 8.78\% with 64 neurons in the hidden layer.

\subsection{SVM-based Classification}\label{ssec:SVM}
A Support Vector Machine (SVM) is a method of supervised learning able to analyze data and recognize patterns. It is here used as a non-probabilistic binary linear classifier. The initial feature space is mapped using a Gaussian kernel so as to maximize the final linear separability between classes. The criterion of good separability is that the hyperplane of decision should have the largest distance to the closest training data points of any class.

Experiments are here performed based on the SVM implementation available in the Torch toolbox. Using 20 features, as it was shown with ANN and GMM to convey almost all the information contained in the large feature set, we obtained the following performance measures: \textbf{TPR=81.87\%, FPR=0.32\% and RER=18.13\%}. SVM is then clearly outperformed by the 2 other classifiers, the GMM approach providing the best identification rates.

\section{Conclusion}\label{sec:Conclu}
This paper focused on the problem of cough detection relying only on audio recordings, as a preliminary and necessary study before integrating other sensors. A large set of features characterizing various aspects of the audio signal was proposed. These features were first assessed based on information theoretical measures, evaluating not only their intrinsic discrimination power, but also their redundancy and complementarity. Secondly, cough detection on recordings from patients suffering from mucoviscidosis was performed with three types of classifier: SVM, ANN and GMM. Impacts of feature dimension (reduced using a mutual information-based feature selection algorithm) as well as of the classifier complexity were analyzed. The best results were obtained with the GMM approach using only 20 features, reporting a sensitivity of 95.2\% and a specificity of 94.3\%.

\section{Acknowledgments}\label{sec:Acknowledgments}

Thomas Drugman is supported by the ``Fonds National de la Recherche Scientifique'' (FNRS). The authors would like to thank the belgian mucoviscidosis center at the Cliniques Universitaires Saint-Luc for providing the audio recordings, as well as Pascaline Delchambre for her preliminary experiments. The authors would like to
thank the Walloon Region, Belgium, for its support (grant WIST 3 COMPTOUX \# 1017071).

\bibliographystyle{unsrt}
\bibliography{refs}

\begin{thebibliography}{10}

\bibitem{CorrelQL}
S.~Decalmer, D.~Webster, A.~Kelsall, K.~McGuinness, A.~Woodcock, and J.~Smith.
\newblock Chronic cough : how do cough reflex sensitivity and subjective
  assessments correlate with objective cough counts during ambulatory
  monitoring?
\newblock In {\em Thorax}, volume~62, pages 329--334, 2007.

\bibitem{Smith}
J.~Smith.
\newblock Cough: Assessment and equipment.
\newblock In {\em The Buyers' Guide to Respiratory Care Products}, pages
  96--101, 2008.

\bibitem{ERS}
A.~Morice, G.~Fontana, M.~Belvisi, S.~Birring, and K.~Chung et~al.
\newblock Ers guidelines on the assessment of cough.
\newblock In {\em European Respiratory Journal}, volume~29, pages 1256--1276,
  2007.

\bibitem{Peeters}
G.~Peeters.
\newblock A large set of audio features for sound description (similarity and
  classification) in the cuidado project.
\newblock 2003.

\bibitem{MFCC}
F.~Zheng, G.~Zhang, and Z.~Song.
\newblock Comparison of different implementations of mfcc.
\newblock In {\em J. Computer Science and Technology}, volume 16(6), 2001.

\bibitem{Shue}
Y.~Shue, G.~Chen, and A.~Alwan.
\newblock On the interdependencies between voice quality, glottal gaps, and
  voice-source related acoustic measures.
\newblock In {\em Interspeech Conference}, volume 34:37, 2010.

\bibitem{DrugmanPhase}
T.~Drugman, T.~Dubuisson, and T.~Dutoit.
\newblock Phase-based information for voice pathology detection.
\newblock In {\em Int. Conf. on Acoustics, Speech and Signal Processing}, 2011.

\bibitem{STRAIGHT_F0}
H.~Kawahara, H.~Katayose, A.~de~Cheveigne, and R.~Patterson.
\newblock Fixed point analysis of frequency to instantaneous frequency mapping
  for accurate estimation of f0 and periodicity.
\newblock In {\em Proc. Eurospeech}, volume~6, pages 2781--2784, 1999.

\bibitem{FSBook}
L.~Huan and H.~Motoda.
\newblock Feature selection for knowledge discovery and data mining.
\newblock In {\em The Springer International Series in Engineering and Computer
  Science}, volume 454, 1998.

\bibitem{Cover}
T.~Cover and J.~Thomas.
\newblock Elements of information theory.
\newblock In {\em Wiley Series in Telecommunications, New York}, 1991.

\bibitem{Drugman-MI}
T.~Drugman, M.~Gurban, and J-P Thiran.
\newblock Relevant feature selection for audio-visual speech recognition.
\newblock In {\em IEEE International Workshop on Multimedia Signal Processing},
  2007.

\bibitem{Torch}
R.~Collobert, S.~Bengio, and J.~Mariéthoz.
\newblock Torch: a modular machine learning software library.
\newblock In {\em Technical Report IDIAP-RR 02-46}, 2002.

\end{thebibliography}

%
%

\end{document}